\documentclass[11pt,preprint]{aastex}

\usepackage{graphics,graphicx}
\usepackage{natbib}
\usepackage{epstopdf}
\newcommand{\gtorder}{\mathrel{\raise.3ex\hbox{$>$}\mkern-14mu
            \lower0.6ex\hbox{$\sim$}}}
\newcommand{\ltorder}{\mathrel{\raise.3ex\hbox{$<$}\mkern-14mu
            \lower0.6ex\hbox{$\sim$}}}

\begin{document}

\title{\small\rm ERRATUM: ``DETERMINING NEUTRON STAR MASSES AND RADII USING ENERGY-RESOLVED WAVEFORMS OF X-RAY BURST OSCILLATIONS''(ApJ, 776, 19 [2013])}

{\small\scshape Ka Ho Lo, M. Coleman Miller, Sudip Bhattacharyya, and Frederick K. Lamb}

\vspace{6pt}

This paper contains two errors: (1)~an incorrect expression was used for the differential emitting area on the stellar surface, as measured in the comoving frame, and (2)~our results were described incorrectly in the caption of Figure 5 and the text on page 19, and incompletely on page 24.

\section{Differential Emitting Area}

The physical properties of the stellar atmosphere that appear in the radiation transport and other equations that would determine the properties of the hot spots considered in this work are defined in a local inertial frame that is momentarily comoving with the rotating atmosphere of the star and naturally yield the sizes, shapes, temperature distributions, and other properties of the hot spots as functions of linear (rather than angular) dimensions in the atmosphere, when measured in the frame comoving with the atmosphere. For example, computations of the evolution of the heated region produced by a thermonuclear X-ray burst yield the linear dimensions of the heated region as a function of time when measured in the comoving frame.

In this work, the star was assumed to be uniformly rotating and spherical, the exterior spacetime was assumed to be the Schwarzschild spacetime, and for simplicity, the dimensions of the phenomenological hot spots that were considered were defined in angular rather than linear coordinates. We describe physics within the stellar atmosphere using a spherical polar coordinate system that corotates with the star and define the colatitude $\theta'$ in this coordinate system as the colatitude $\theta$ in the Schwarzschild coordinate system. Angular separations are unchanged when one transforms from the rotating frame to the static frame of the Schwarzschild spacetime, so the angular separations $d\theta$ and $d\phi$ of two points on the stellar surface when measured in the Schwarzschild coordinate system are the same as the angular separations $d\theta'$ and $d\phi'$ of the same two points on the stellar surface when measured in the rotating frame.

In a differential time $dt'$, a burning or heating front that advances with speed $v_f$ in the local comoving frame will cover a linear differential distance $dx' = v_f dt'$ as measured in the comoving frame.
The expression for the differential area $dS'$ defined by the locally orthogonal differential \textit{linear} coordinate intervals $dx'$ and $dy'$ on the stellar surface is simply $dS'= dx'dy'$ when measured in the rotating frame by observers who use centrally synchronized clocks (see \S~IV of \citealt{2012AmJPh..80..772K}). In contrast, the expression for the differential surface area $dS'$ defined by the differential \textit{angular} coordinate intervals $d\theta'$ and $d\phi'$ is $dS' = R^2 \sin\theta' d\theta' d\phi' \gamma(\theta')$ when measured in the rotating frame (again see \S~IV of \citealt{2012AmJPh..80..772K}). Here $R$ is the radius of the star in the corotating frame and $\gamma(\theta') \equiv 1/[1 - (v_\phi(\theta')/c)^2]^{1/2}$ is the Lorentz gamma factor for the linear azimuthal speed $v_\phi(\theta')$ of the gas in the atmosphere at colatitude $\theta'$ as measured in the static frame, where $v_\phi(\theta') = \Omega' R\sin\theta'$ in terms of the angular velocity $\Omega'$ of the star measured at $R$ and $\Omega' = \Omega (1 - 2GM/Rc^2)^{-1/2}$ in terms of the angular velocity $\Omega$ of the star measured at infinity.

The reason for the factor $\gamma(\theta')$ in the expression for $dS'$ when $dS'$ is expressed in angular coordinates is that the differential \textit{linear} interval that corresponds to the differential \textit{angular} interval $d\phi'$ at colatitude $\theta'$ is $R\sin\theta' d\phi' \gamma(\theta')$ when measured by observers in the rotating frame who use centrally synchronized clocks. As a result, the distance $L'$ around the star at a constant colatitude $\theta'$ is $2\pi R\sin\theta'\gamma(\theta')$ when measured by such observers. When it is measured by observers in the local static (Schwarzschild) frame,  the differential linear interval that corresponds to the differential angular interval $d\phi$ is Lorentz-contracted by the factor $1/\gamma(\theta)$ and is therefore $R \sin\theta d\phi$ (the radius of the star measured in the static frame is the same as the radius measured in the rotating frame). The distance $L$ around the star at a constant colatitude $\theta$ is therefore $L'/\gamma(\theta) = 2\pi R \sin\theta$ when measured by observers in the static frame.

Although the radiation transport equation used in \cite{2013ApJ...776...19L} was correct, the factor $\gamma(\theta)$ was mistakenly omitted from the expression for the azimuthal dimension of the emitting area when measured in the rotating frame. This caused the computed observed flux from each differential emitting area to be too small by a factor of $1/\gamma(\theta')$. We found and corrected this error in our codes in September 2016, after being told of discrepancies between pulse waveforms given by our codes, which follow the radiation forward in time from the star to the observer, and the corresponding waveforms given by the codes developed by Psaltis and collaborators (see, e.g., \citealt{2012ApJ...745....1P}) and by \cite{2017arXiv170907292N}, which follow the radiation backward in time from the observer to the star, and after discussing these discrepancies with N{\"a}ttil{\"a} and Poutanen and several members of the \textit{NICER} (Neutron Star Interior Composition Explorer) Pulse Waveform Modeling and Analysis Working Group.

This error is present not only in all our previously published computations of pulse waveforms (e.g., \citealt{mill98a, lamb09a, lamb09b, 2013ApJ...776...19L}; and \citealt{2015ApJ...808...31M}), but also in all other previously reported computations of absolute pulse waveform fluxes that followed the radiation forward in time from the star to the observer (e.g., \citealt{pout03, pout04, cade07, 2007ApJ...663.1244M, leah08, leah09}; and \citealt{2011ApJ...726...56M}). (All the codes that followed the radiation from the star to the observer had been verified by cross-comparing their results for hot spots defined in angular coordinates, for a large number of cases.) 
We note that the idealized radiation spectra, beaming patterns, and spot shapes assumed in many of the previously reported computations of pulse waveforms very likely introduced systematic errors that are much larger than the errors introduced by using a differential emitting area that was too small by a factor of $1/\gamma(\theta')$.
This error has now been corrected in all our codes, including the ones we are using to analyze \textit{NICER} data, and it is our understanding that it has also now been corrected in all the other codes that follow the radiation from the star to the observer. We have compared results produced by our corrected codes with the results produced by other corrected codes and have verified that the results agree to very high accuracy (e.g., the pulse waveforms produced by the corrected codes differ by 0.1\% or less at all phases and energies).

Linear distances in the east-west direction at colatitude $\theta$ are a factor of $1/\gamma(\theta)$ smaller when measured by observers in the local static frame than when measured by observers rotating with the star. This means that a hot spot that is circular when measured in linear distances in the frame comoving with the atmosphere (for example, because it is produced by transport of heat at the same speed in all directions from a small, very hot initial spot, as measured in the frame comoving with the atmosphere) will appear oval (narrower in the east-west direction than in the north-south direction) when measured in linear distances in the static frame, or in angular distances in either frame. Conversely, a spot that is circular in its angular dimensions will appear circular when measured in linear dimensions in the static frame, but will appear oval (in this case, wider in the east-west direction than in the north-south direction) when measured in linear dimensions in the rotating frame. These effects must be taken into account when determining how a spot, and the waveform it produces, appear to a distant observer.

When the omission of the factor $\gamma(\theta)$ in our expression for the differential area $dS'$ of the emitting surface defined by differential angular intervals is corrected, the flux from each such differential area is increased by the factor $\gamma(\theta)$. The effect of this increase in the flux on the pulse waveform depends on the size and location of the hot spot, and the radius and rotation frequency of the star. 

For spots that were very small, omission of the factor $\gamma(\theta)$ caused the computed \textit{amplitude} of the waveform to be a factor $1/\gamma(\theta_s)$ smaller than it should have been, where $\theta_s$ is the colatitude of the spot center, but it did not affect the \textit{shape} of the waveform. The fractional error $\gamma(\theta_s) - 1$ in the computed amplitude of a very small spot was smaller for spots nearer the spin axis, because $v_\phi(\theta_s)$ is smaller and $\gamma(\theta_s)$ is closer to unity there. The amplitude of the waveform has no practical significance if (a)~the uncertainty in the distance to the pulsar is much larger than the error that was made in computing the amplitude of the waveform or (b)~the absolute amplitude of the waveform is unimportant for the application at hand, for other reasons. However, for some applications, such as estimating the mass and radius of a star that has a very accurately known distance, the absolute amplitude of the waveform may be important.

For spots that were not small, omission of the factor $\gamma(\theta)$ distorted the computed \textit{shape} of the waveform, as well as its amplitude. The value of $\gamma(\theta) - 1$ at the point on the spot farthest from the rotational pole places an upper bound on the maximum fractional error in the waveform flux at any phase. The actual maximum fractional error in the waveform flux at any phase was typically much smaller than this, because the flux observed at a given pulse phase is a sum of fluxes from different colatitudes. The error in the waveform shape was smaller still, because it is determined by the appropriately weighted variation of $\gamma - 1$ over the entire spot, which was usually much smaller. The distortion of the waveform shape was less for waveforms that were produced by smaller spots (as noted above, the waveforms of very small spots were not distorted). The distortion of the waveform produced by a spot was also less if its center was near the rotational pole, because $\gamma(\theta) - 1$ vanishes at the rotational pole, or if it was near the rotational equator, because the variation in $\gamma(\theta) - 1$ vanishes at the rotational equator. 

All the synthetic waveform data we analyzed in \cite{2013ApJ...776...19L} were generated assuming a stellar mass of $M=1.6~M_\odot$ and a stellar radius of $R=5GM/c^2=11.813$~km, and almost all were generated assuming a circular hot spot with an angular radius of 25$^\circ$. Our ``low-inclination'' reference case assumed a spot colatitude of 20$^\circ$ and a stellar spin rate of 400~Hz, whereas our ``high-inclination'' reference case assumed a spot colatitude of 90$^\circ$ and a stellar spin frequency of 600~Hz. For these spot geometries, the azimuthal speed, and hence the Lorentz $\gamma$ factor, is greatest at the point on the spot that is farthest from the rotational pole. In the low-inclination case, this point is at the lowest-colatitude edge of the spot, where $v_{\phi,\rm{max}} = 2.710\times 10^9$~cm~s$^{-1}$ and $v_{\phi,\rm{max}}/c = 0.09041$; at this point, $\gamma = 1.00411$. The fractional errors in the waveform fluxes quoted for this case were therefore $\le 0.4$\% and were usually much less than this. In the high-inclination case, the highest azimuthal speed within the spot occurs at the center of the spot, where $v_\phi = 5.749\times 10^9$~cm~s$^{-1}$ and $v_{\phi,\rm{max}}/c = 0.1918$; at this point, $\gamma = 1.0189$. The fractional errors in the waveform fluxes quoted for this case were therefore less than 2\% and were usually much less than this. The distortions in the reported shapes of the waveforms produced by the circular spots with an angular radius of 25$^\circ$ were smaller still.

In \cite{2013ApJ...776...19L} and \cite{2015ApJ...808...31M}, the synthetic waveform data and the model waveforms that were fit to these data were affected by the error in the emitting area in exactly the same way. Consequently, it is unlikely that this error significantly affected our estimates of $M$ and $R$, their statistical uncertainties, or the systematic errors in these estimates.

\section{Knowledge of the Background}

The results listed in Table~2 and shown in the figures in \cite{2013ApJ...776...19L} are described correctly, but the caption of Figure~5, the final paragraph in the text on page~19, and point~2 in the middle of the left column on page 24 are incorrect or incomplete and therefore misleading.

The caption of Figure 5 says that $M$ and $R$ are much more tightly constrained if the background is known. This is not generally true. It can be true if the geometry is poor and the constraints are weak (compare, e.g., Figure~5(b) with Figure~2(d)), but if the geometry is favorable and the constraints are tight, independent knowledge of the background makes little difference (compare, e.g., Figure~5(a) with Figure~2(c)).

The text on page~19 says that ``knowing the size and spectrum of the background greatly improves the constraints on $M$ and $R$. For our high-inclination case, knowing the background decreases the 1$\sigma$ uncertainties in $M$ and $R$ from $\sim\,$9\% to $\sim\,$4\% and the 3$\sigma$ uncertainties from $\sim\,$50\% to $\sim\,$9\% (compare Figure~5(a) with Figure~2(e)). For our low-inclination case, knowing the background leads to a 1$\sigma$ uncertainty of $\sim\,$20\%, whereas without this knowledge one obtains no useful constraints on $M$ and $R$ (compare Figure~5(b) with Figure~2(f)).'' These statements are incorrect. Figure~5(a) should be compared with Figure~2(c), not Figure~2(e), and Figure~5(b) should be compared with Figure~2(d), not Figure~2(f). When the correct comparisons are made, one again concludes that knowledge of the background can improve the constraints if the geometry is poor and the constraints are therefore weak, but if the geometry is favorable and the constraints are therefore tight, independent knowledge of the background makes little difference.

Point~1 in the second group of numbered points in the middle of the left column on page~24 states that ``independent information about the background can greatly reduce the uncertainties in estimates of $M$ and $R$''. Again, this can be true if the geometry is poor and the constraints are weak, but this statement is incomplete and is therefore potentially misleading.

We thank Slavko Bogdanov, Simin Mahmoodifar, Sharon Morsink, Joonas N{\"a}ttil{\"a}, Feryal Ozel, Juri Poutanen, Dimitrios Psaltis, and Tod Strohmayer for useful discussions.

\bibliography{lo-erratum}
\end{document}